\documentclass[amsmath,amssymb,aps,pra,10pt,twocolumn]{revtex4-1}

\usepackage{graphicx}
\usepackage{dcolumn}

\begin{document}

\title{Structure and thermochemistry of K${}_2$Rb, KRb${}_2$ and K${}_2$Rb${}_2$}

\author{Jason N. Byrd}
\author{John A. Montgomery, Jr.}
\author{Robin {C\^ot\'e}}
\affiliation{Department of Physics, University of Connecticut, Storrs, CT 06269}

\begin{abstract}
The formation and interaction of ultracold polar molecules is a topic of
active research.  Understanding possible reaction paths and molecular
combinations requires accurate studies of the fragment and product energetics.
We have calculated accurate gradient optimized ground state structures and zero point
corrected atomization energies for the trimers and tetramers formed by the
reaction of KRb with KRb and corresponding isolated atoms.  The K${}_2$Rb and
KRb${}_2$ trimers are found to have global minima at the $C_{2v}$ configuration
with atomization energies of $6065$ and $5931$ cm${}^{-1}$ while the
tetramer is found to have two stable planar structures, of $D_{2h}$ and $C_s$
symmetry, which have atomization energies of $11131$ cm${}^{-1}$ and
$11133$ cm${}^{-1}$, respectively.   We have calculated the minimum energy
reaction path for the reaction KRb$+$KRb to K$_2+$Rb$_2$ and found it to be
barrierless.
\end{abstract}

\maketitle

The formation and interaction of
ultracold polar molecules is a topic of great current interest in
physics.  New techniques for the formation of rovibrational ground state polar
molecules via STIRAP\cite{vitanov2001} (stimulated rapid adiabatic passage) or
FOPA\cite{philippe2008} (Feshbach-optimized photo-association) allow
experiments to be performed with $v=0$ heteronuclear diatomic molecules,
including KRb\cite{zirbel2008,ospaelkaus2008,ni2008,opelkaus2010} and LiCs\cite{deiglmayr2008}
Proposals for quantum computation
with polar molecules\cite{yelin2006,kuznetsova2008} have generated a growing need
for understanding of the dynamics of diatom-diatom collisions.  Such studies of
diatomic dynamics require knowledge of the open and closed channels relevant in
those reactions. The purpose of the present paper is to present accurate
{\em ab initio} calculations of the structure and thermochemistry of 
several chemical species relevant to the study of KRb$-$KRb dimer interactions.

Theoretical work on electronic structure of few-body alkali systems has been limited to
lighter homonuclear trimers, in particular doublet \cite{byrd2009-a} and
quartet \cite{cvitas2007} Li${}_3$, doublet K${}_3$ \cite{hauser2008} and
quartet Na${}_3$ \cite{simoni2009}.  The recent work of \.{Z}uchowski and
Hutson\cite{zuchowski2010} has characterized the atomization energy of the
alkali homo- and heteronuclear triatomic species formed from Li, Na, K, Rb, and Cs. 
These homonuclear trimers have $A'$
ground electronic states in $C_s$ symmetry that correlate to $B_2$
symmetry in $C_{2v}$.  Previous mixed alkali tetramer studies have been limited to
structure studies of Li${}_n$X${}_m$ (X$=$Na and K)
\cite{dahlseid1992,jiang2006} and that of RbCs$+$RbCs \cite{tscherbul2008}.
To date no such calculations have been reported for the heteronuclear
K${}_n$Rb${}_m$ tetramer molecules.

Electronic structure calculations were performed on
K$_2$, Rb$_2$, KRb, K$_2$Rb, KRb$_2$, and K$_2$Rb$_2$
at the CCSD(T) \cite{purvis1982}
level of theory.
As core-valence effects can be important in alkali metals, we correlate the inner valence
electrons in potassium, keeping only 1s$^2$2s$^2$2p$^2$ in the core.  Rubidium is heavy enough
that relativistic effects are significant, so we replace its inner shell electrons
by the Stuttgart small-core relativistic (ECP28MDF) ECP \cite{lim2005}.
Basis sets are taken from the Karlsruhe
def2-TZVPP \cite{weigend2005} and def2-QZVPP \cite{weigend2003} orbital
and fitting sets.

Optimized geometries for
K$_2$, Rb$_2$, KRb, K$_2$Rb, KRb$_2$, and K$_2$Rb$_2$
were found at the CCSD(T)/def2-TZVPP level of theory.
Calculation of the harmonic vibrational frequencies was done
to verify that the calculated structures were minima on the
potential energy surface, and the calculated frequencies 
were used to obtain vibrational zero point energy (ZPE) corrections. 
These structures were further optimized at the CCSD(T)/def2-QZVPP level
of theory, leading to a $0.07$~\AA~ correction in the
bond lengths and $~60$~cm${}^{-1}$ in final atomization energies.  The
CCSD(T)/def2-QZVPP geometries are tabulated in Table \ref{geom}.  

\begin{table*}
\caption{\label{geom} Calculated CCSD(T)/QZVPP molecular
        geometries (in Angstroms and degrees).}
\begin{ruledtabular}
\begin{tabular}{lccccc}
 &
\multicolumn{1}{c}{$r_e$} & & & &\\
K${}_2$           & 3.956  & & & &\\
Rb${}_2$          & 4.233  & & & &\\
KRb               & 4.160  & & & &\\
\hline
\vspace{0.1cm}
&
\multicolumn{1}{c}{$r_{\rm K-Rb}$}   &
\multicolumn{1}{c}{$r_{\rm K-Rb}'$}   &
\multicolumn{1}{c}{$\theta$} & & \\
K${}_2$Rb $C_{2v}$& 4.279  & 4.279  & 70.68  & &  \\
K${}_2$Rb $C_s$   & 4.361  & 5.234  & 48.81   & & \\
KRb${}_2$ $C_{2v}$& 4.271  & 4.271  & 82.13   & & \\
KRb${}_2$ $C_s$   & 4.193  & 5.179  & 57.07   & & \\
\hline
 &
\multicolumn{1}{c}{$r_{\rm Rb-Rb}$}   &
\multicolumn{1}{c}{$r_{\rm K-K}$}   &
\multicolumn{1}{c}{$r_{\rm K-Rb}$}   &
\multicolumn{1}{c}{$\theta_{\rm K-Rb-Rb}$}   &
\multicolumn{1}{c}{$\theta_{\rm K-K-Rb}$} \\
K${}_2$Rb${}_2$ $D_{2h}$  & 8.224  & 4.0307 & 4.579 & &   \\
K${}_2$Rb${}_2$ $C_s$     & 4.761  & 4.408  & 4.189 & 53.34 & 55.476 \\
\end{tabular}
\end{ruledtabular}
\end{table*}

\begin{table*}
\caption{\label{spec} Dissociation and zero point energies calculated using CCSD(T) and
CCSD(T)-F12b correlation methods with successive basis sets and CBS
extrapolated
values (in cm${}^{-1}$).}
\begin{ruledtabular}
\begin{tabular}{lddddddd}
 &
\multicolumn{1}{c}{ZPE TZVPP} &
\multicolumn{2}{c}{$D_e$ TZVPP} &
\multicolumn{2}{c}{$D_e$ QZVPP} &
\multicolumn{2}{c}{$D_0$ CBS}\\
 &
\multicolumn{1}{c}{CCSD(T)} &
\multicolumn{1}{c}{CCSD(T)} & \multicolumn{1}{c}{CCSD(T)-F12b} &
\multicolumn{1}{c}{CCSD(T)} & \multicolumn{1}{c}{CCSD(T)-F12b} &
\multicolumn{1}{c}{CCSD(T)} & \multicolumn{1}{c}{CCSD(T)-F12b}\\
\hline
K${}_2$\footnote{Experimental value $4405.389$ cm${}^{-1}$\cite{falke2006}.}
& 46.0 & 4098.8  & 4276.9  & 4460.0  & 4369.7  & 4677.6  & 4391.5 \\
Rb${}_2$\footnote{Experimental value $3965.8$ cm${}^{-1}$\cite{amiot1990}.}
& 26.8 & 3494.3  & 3723.3  & 3842.7  & 3885.4  & 4070.2  & 3976.8 \\
KRb\footnote{Experimental value $4180.417$ cm${}^{-1}$\cite{ni2008}.}
& 35.4 & 3829.4  & 4015.6  & 4135.6  & 4128.7  & 4323.6  & 4175.7 \\
\hline
K${}_2$Rb $C_{2v}$
& 69.8  & 5588.2  & 5805.5  & 6067.7  & 5995.7  & 6574.2  & 6009.4 \\

K${}_2$Rb $C_s$
& 72.4  & 5606.3  & 5843.7  & 6179.1  & 6015.9  & 6524.7  & 6069.1\\

KRb${}_2$ $C_{2v}$
& 62.8  & 5394.5  & 5635.1  & 5911.0  & 5842.2  & 6043.5  & 5788.3 \\

KRb${}_2$ $C_s$
& 59.0  & 5215.9  & 5475.4  & 5728.5  & 5690.4  & 6225.1  & 5930.5 \\
\hline
K${}_2$Rb${}_2$ $D_{2h}$
& 129.5 & 10210.8  & 10669.4  & 11275.3  & 11011.1  & 11922.7  & 11131.0 \\

K${}_2$Rb${}_2$ $C_s$
& 126.2 & 10198.3 & 10629.9 & 11211.4 & 10993.7 & 11824.6 & 11133.0
\end{tabular}
\end{ruledtabular}
\end{table*}

Evaluation of the contribution of scalar relativistic corrections to K${}_2$
indicate a small $0.005$~\AA~ and $<8$~cm${}^{-1}$ contribution in all electron
correlation calculations\cite{iron2003}, while for Rb${}_2$ it has been
shown \cite{lim2005b} that the small core Stuttgart pseudopotential 
gives an accurate representation of relativistic effects on the bond length
and dissociation energy.

Single point energy calculations were then done using the CCSD(T)-F12b \cite{adler2007,knizia2009}
(explicitly correlated CCSD(T)) level of theory.  The use of explicitly
correlated methods accelerate the slow convergence of the one-particle basis set
by including terms containing the inter-electron coordinates into the wavefunction
\cite{helgaker2008}, thus yielding very
accurate results using triple and quadruple zeta basis sets.
In addition,
we estimate the complete basis set (CBS) limit using the two-point
extrapolation formula of Helgaker {\it et al} \cite{helgaker1997}
\begin{equation}
E_{\rm CBS}=\frac{n^3 E_n - (n-1)^3 E_{n-1}}{n^3-(n-1)^3}.
\end{equation}
In Table \ref{spec} the CCSD(T) and CCSD(T)-F12b dissociation energies
for the def2-TZVPP and def2-QZVPP basis sets are tabulated as well as the zero point
energy (ZPE) corrected atomization energies.  After extrapolation, the diatomic
CCSD(T)-F12b ZPE corrected dissociation energies agree very well with the
experimental diatomic dissociation energies, as shown in Table \ref{spec}.
The {\it ab initio} calculations were done
using the Gaussian 09 \cite{gaussian09short} and MOLPRO 
\cite{molpro09short,hampel1992,knowles1993} packages.

We have found that both K${}_2$Rb and KRb${}_2$ have two energetically close
local minima on the ground state surface, one of $C_{2v}$ symmetry
and another less symmetric $C_s$ structure (geometries given in Table \ref{geom}).
While dependent on the level of theory used to evaluate the atomization energy,
we conclude that the symmetric $C_{2v}$ geometry is the global minima for each
trimer.  The atomization energies calculated are found to be in good agreement
with those recently published by \.{Z}uchowski and Hutson \cite{zuchowski2010}.

The K$_2$Rb$_2$ tetramer is found to have two nearly degenerate minima on the
potential energy surface.  One is a rhombic structure of $D_{2h}$ symmetry,
and another planar ($C_s$) structure that corresponds to an interchange of K
and Rb atoms. These structures are bound by $\sim 3000$ cm$^{-1}$ with respect to
K$_2$+Rb$_2$ or KRb+KRb.
The electronic structure of these two isomers is very similar,
and their stability is likely due to three-center bonds of the sort
proposed for Li$_n$Na$_{4-n}$ clusters\cite{dahlseid1992,jiang2006}.
The rhombic K$_2$Rb$_2$ structure has a short
($\sim 4$\AA) distance and a long ($\sim 8$\AA) Rb-Rb distance.
The equivalent structure where the K-K distance is short and the Rb-Rb
distance is long is found to be a transition state, not a stable minimum.

To determine if there is any barrier to the
KRb$+$KRb$\rightarrow$K$_2$Rb$_2\rightarrow$Rb${}_2+$K${}_2$ reaction, we
calculate a minimum energy path for the KRb$+$KRb$\rightarrow$K$_2$Rb$_2$ and
Rb${}_2+$K${}_2$$\rightarrow$K$_2$Rb$_2$ reactions.  We start by locating the
minimum energy geometric configuration at long range.
This is done by calculating {\it ab initio} the dipole and
quadrupole electrostatic moments of K$_2$, Rb$_2$ and KRb and then minimizing the
long range electrostatic interaction energy \cite{stone} with respect to the
angular configuration of the molecules.  This minimization resulted in a T type geometry
for both K$_2+$Rb$_2$ and KRb$+$KRb.  We have recently shown that long-range expansions
of this type accurately reproduce diatom-diatom interaction energies \cite{zemke2010}.  
From these initial geometries, the reaction path was followed by freezing the
diatom-diatom distance and optimizing the diatomic bond lengths and angular
orientations at the frozen core CCSD(T)/def2-TZVPP level of theory.
Single point energies were evaluated along this path using the CCSD(T)-F12b
level of theory including the core-valence correlation energy and extrapolated
to the CBS limit as discussed above. 
This procedure, in which a high level energy profile is evaluated
along a reaction path calculated at a lower level of theory, is known to
be a good approximation to the energy profile along the reaction path calculated
at the high level of theory \cite{malick1998}.

\begin{figure}
\resizebox{8.5cm}{!}{\includegraphics{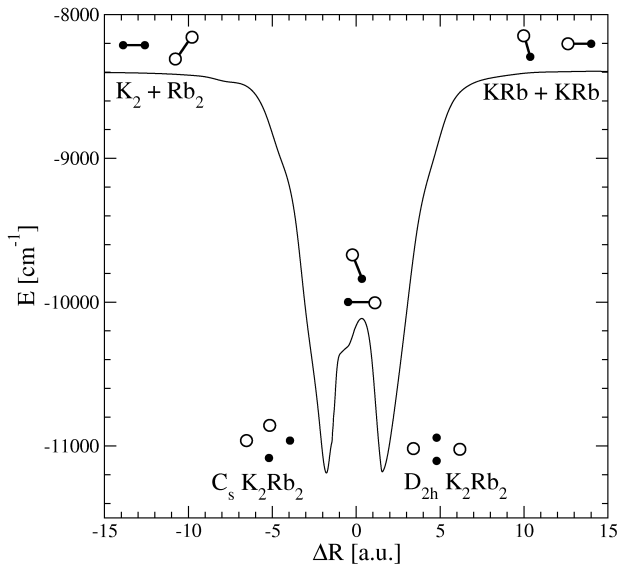}}
\caption{\label{reactionpath} Minimum energy path connecting the KRb$+$KRb and
K${}_2+$Rb${}_2$ dissociation limits.  Included are schematic geometric at
points of interest, where open and closed circles represent rubidium and
potassium atoms respectively.}
\end{figure}

We find that
the KRb$+$KRb dissociation limit connects to the $D_{2h}$ minima while the
K$_2+$Rb$_2$ dissociation limit connects to the $C_{s}$ minima, with no barrier
found to either reaction.
A similar conclusion was obtained for the
RbCs$+$Rbcs$\rightarrow$Rb$_2+$Cs$_2$ reaction by Tscherbul {\it et
al}\cite{tscherbul2008}.
To finish characterizing the reaction path going from
dissociation limit to the other, we locate the transition state and calculate
the intrinsic reaction coordinate (IRC) \cite{fukui1981} reaction path
connecting the $C_s$ and $D_{2h}$
minima structures at the same level of theory as describe above.  
Optimizing the transition state geometry at the inner valence CCSD(T)/def2-TZVPP
discussed previously and evaluating an accurate atomization energy using our 
CCSD(T)-f12b prescription we find that
the transition state is $1167.3$ cm${}^{-1}$ above the $D_{2h}$ dissociation
energy.
The calculated reaction path is plotted in Figure \ref{reactionpath} using the
approximate reaction coordinate
\begin{equation}
\Delta R = (R_{\rm Rb-Rb} + R_{\rm Cs-Cs})/2 - (R_{\rm Rb-Cs} + R'_{\rm
Rb-Cs})/2
\end{equation}
where $R_{\rm A-B}$ is the
distance between atoms A and B.

\begin{figure}
\resizebox{8.5cm}{!}{\includegraphics{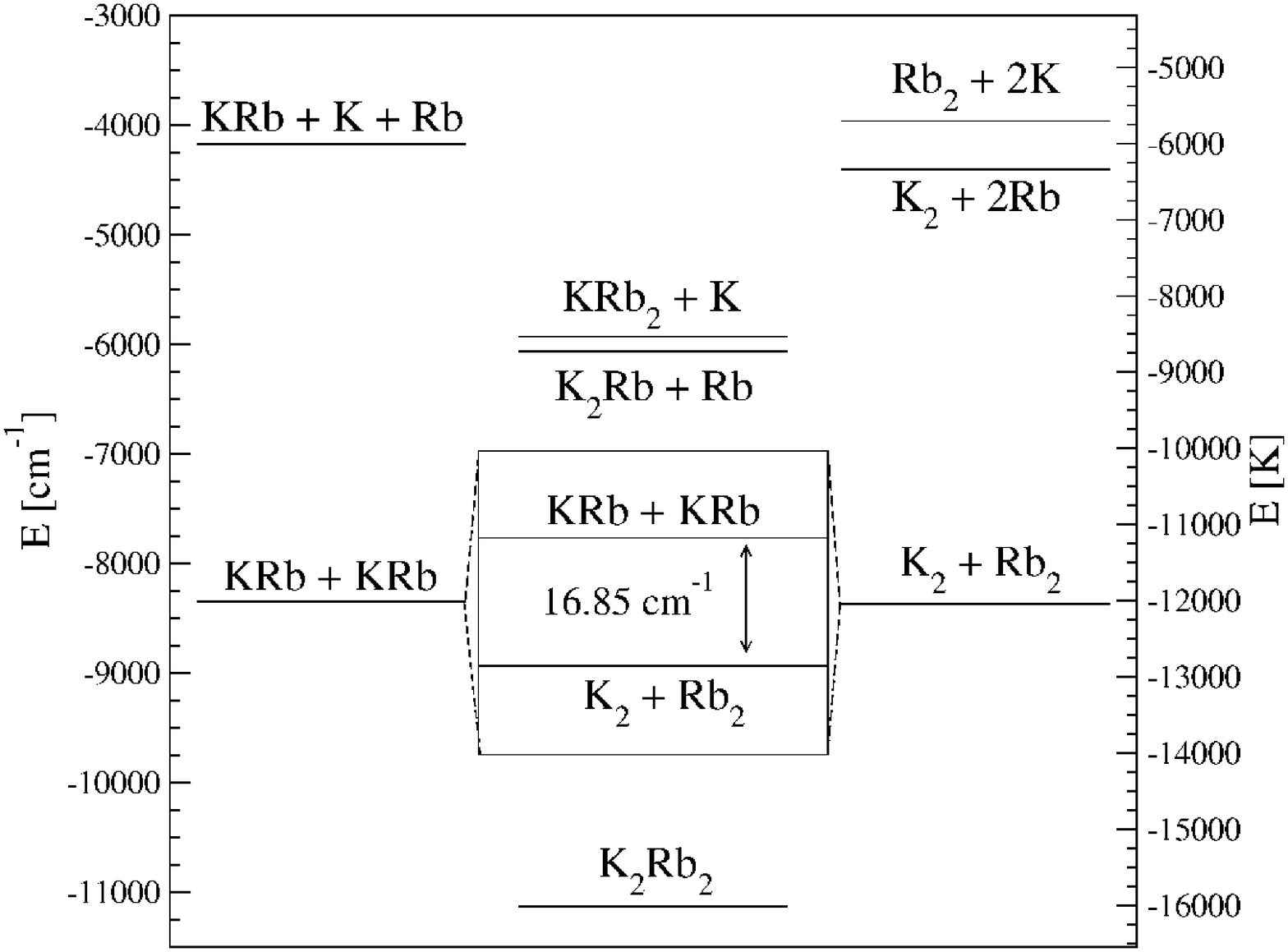}}
\caption{\label{reaction} Schematic energy level diagram for fragment and
structure energies involving KRb with KRb and separated atoms.  Inset figure
shows the small difference between the KRb$+$KRb and K$_2+$Rb$_2$ asymptotes.}
\end{figure}

The formation and trapping of rovibrational ground state KRb diatoms with a high
phase space density\cite{ni2008} offers the opportunity to study chemical
reactions in the ultra-cold regime\cite{opelkaus2010}.  As seen in Figure \ref{reaction}, the
three-body reaction KRb$+$Rb$\rightarrow$Rb${}_2+$K is energetically forbidden
at ultra-cold temperatures, leaving the endothermic four-body reaction
KRb$+$KRb$\rightarrow$Rb${}_2+$K${}_2$ as the only pathway to forming Rb${}_2$
within the trap.  Measurements of the population of Rb${}_2$ within the trap
will then allow direct probing of the exchange reaction rate of KRb$+$KRb.
Inherent in this exchange reaction is the possibility of studying the role of
fermionic/bosonic spin statistics in ultra-cold chemical
reactions\cite{cvitas2005a,cvitas2005b,dincao2008,dincao2009,petrov2005,hudson2008,marcelis2008}.
In this temperature regime, $s$-wave scattering of fermionic ${}^{40}$KRb is
suppressed which should greatly diminish the reaction rate of
${}^{40}$KRb$+$${}^{40}$KRb, thus leaving the trap stable to four-body losses.
If instead the trap was formed with bosonic ${}^{39}$KRb or ${}^{41}$KRb
molecules, no such collisional suppression is expected, where we then expect
comparably large reaction rates to occur.  It is also possible to explore
recent theoretical predictions\cite{marcelis2008} which show that if a bosonic dimer is
composed of two fermions of very different masses the resulting exchange
reaction should still be suppressed despite the overall bosonic nature.  This
could be accomplished by using fermionic ${}^{40}$K and a long lived ${}^{84}$Rb
or ${}^{86}$Rb.  The comparison between reaction rates in the above described
interactions can then be used to directly study the effects of fermion/boson
spin statistics to that of chemical reactions.

\begin{acknowledgments}
We would like to thank W. C. Stwalley for helpful discussions during the course of this
work.  JNB would like to thank the U.S. Department of Energy Office of Basic
Energy Sciences for support and RC would like to thank the Department of Defense
AFOSR for partial support.
\end{acknowledgments}


%

\end{document}